\documentstyle[osa,manuscript,epsbox]{revtex}  % DON'T CHANGE %

\begin{document}                % INITIALIZE - DONT CHANGE % %  %

\title{Effective Dimension Transition of the Dynamics of Granular Materials
in the Pipe.}

\author{Akinori Awazu\footnote{E-mail: awa@zenon.ms.osakafu-u.ac.jp}\\ }

\address{Department of Mathematical Sciences\\
Osaka Prefecture University, Sakai 599-8531 Japan.         } %

\maketitle
\begin{abstract}
Phase transition of strongly excited granular materials in 2D 
pipe is investigated numerically. By changing the ratio between width of 
the pipe and the height of the granular bed, we observe the transition 
between the 1 dimensional like state and the actual 2 dimensional state. 
Moreover, it is found that the character of the transition changes 
when the magnitude of dissipation passes through a critical value.\\
\vspace{2mm}
PACS number(s):\\

\end{abstract}

Granular materials exhibit some complex phenomena\cite{re1,re2}. These complex
phenomena
strongly depend on not only the character of external forces but also
that of boundary condition.
When granular materials in slim pipes with fluid (air, water, 
and so on) flow down or are fluidized by inducing fluid from the 
bottom, slugging is observed\cite{tr9,tr12} . In most of the 
investigations of such slugging in a slim pipe, the system's 
effective dimension is regarded as 1\cite{tr10,tr11,tr12}.
On the other hand, two or more dimensional phenomena, bubbling and channeling, 
occur when granular materials are in a wider vessel\cite{fh1}.
The similar tendency is also observed for granular materials in a vibrating
vessel\cite{con0,con00,con1,con2}. When the width of the vessel is small,
most of particles construct solid structure in bulk and only particles near side
walls flow down to the bottom\cite{re1,con00}. However, in a wide
vessel, most of particles are strongly fluidized and construct multi-roll
structure like Bernard convection\cite{con2} From these facts, the
behaviors of granular materials are extremely different between that
in a slim vessel and that in a wide vessel.
Then, an important problems appear; what quantity determines
'slim' (one dimensional system)
or 'wide' (two or three dimensional system) for a given system?
The purpose of this paper is to make a clear view on this problem. 

Here, we simulate following simplified situation\cite{setu}.
The system consists of two dimensional particles of mass 1 
and diameter $d$ in a two-dimensional box under uniform gravity. 
The width of the box is $a$, and the height of the box is 
infinite. We employ the following particle model which is one of the 
simplest 
model of granular materials. The equation of the motion of the $i$th 
particle is
\begin{equation}
\ddot{{\bf x}_{i}}=-\sum_{j=1}^{N}\theta(d-|{\bf x}_{i}-{\bf x}_{j}|)\{{\bf \nabla}U(d-|{\bf x}_{i}-{\bf x}_{j}|)+\eta({\bf v}_{i}-{\bf v}_{j})\}+{\bf g}
\end{equation}
\begin{equation}
U(d-|{\bf x}_{i}-{\bf x}_{j}|)=\frac{k}{2}(d-|{\bf x}_{i}-{\bf x}_{j}|)^{2}
\end{equation}
here, $\theta$ is Heviside function, $N$ is the total number of particles,
$k$ and $\eta$ are respectively the elastic constant and the viscosity 
coefficient\cite{keta}, 
and ${\bf x}_{i}(x_{i},y_{i})$, ${\bf v}_{i}(v_{x_{i}}, v_{y{i}})$,
and ${\bf g}=(0,-g)$ are, respectively, the position, the velocity and 
the gravity of $i$th particles. In this model, the effect 
of particles' rotation is neglected.
The system is driven by a simple energy source at the bottom of the box; 
a particle hitting the bottom with velocity $(v_{x}, v_{y})$ bounces back 
with the velocity $(v_{x}, V (V>0))$. We regard the bottom of box as $x$ 
axis ($y=0$), and $x=0$ as the center of box. The side walls of the
box are put along $x=a/2$ and $x=-a/2$, and the viscosity which works
between these walls and particles is zero.
At the initial condition, we put particles bed with height $b$ on the
bottom of box.
We simulate this system with some combinations of parameters 
$(\eta, a, b)$, where $a$ and $b$ are enough large compared to the 
particle's diameter $d$. The above equations are calculated with the 
Euler's scheme. The time step $\delta t$ is set enough small such 
that $\delta x$, the displacement of the $i$th particle during 
$\delta t$, does not exceed a given value. In this paper, we set 
$(-g,V)$ so that the average height of the center 
of mass of the system $CM=<(\sum y_{i})/N>_{t}$ keeps enough large
compared to $b$.

Figure 1 shows typical snapshots of the system for respectively, (a)
$a<<b$ with $\eta>\eta_{*}$, (b) $a>>b$ with $\eta>\eta_{*}$, (c)
$a<<b$ with $\eta<\eta_{*}$, and (d) $a>>b$ with $\eta<\eta_{*}$.
Here, we fixed $b$, whereas $(-g,V)$ of (a) is same as that of (b), and
$(-g,V)$ of (c) is same as that of (d). 
In cases of (a) and (c), only the particles distribution in the
horizontal direction is symmetric, and the center of mass of this system
moves a little only in the vertical direction.
On the other hand, the particles distribution is non-uniform in vertical and 
horizontal direction, and one convection appears in cases of (b) and (d). 
In order to characterize the system, we introduce the order parameter 
$L(t)=\sum_{i}|x_{i}(t)v_{y{i}}(t)|/N$ which indicates the strength of the 
convection 
of the system. Figure 2 (a) and (b) are typical probability distributions of 
$L(t)$ which is given by one time series for respectively the cases of
$a<<b$ and those of $a>>b$. In Fig.2 (a), the peak of the probability
distribution  of $L(t)$ appears at $L(t)=0$ which means there are no
convection for  the case $a<<b$. In these cases, the effective dimension
of this system can be regarded as 1. We name such states as the
'1D state'.
On the contrary the peak of the probability 
distribution of $L(t)$ appears at $L(t)>0$ in Fig.2 (b) which means a 
convection with finite magnitude appears in the system.
In these cases, the system is actually 2 dimensional system. We name
such states with a convention as the '2D state'.
Such a transition between the 1D state and the 
2D state which depends on the relation between $a$ 
and $b$ is observed in a simple system.  

In our simulation, particles are strongly excited by the energy
source at the bottom of the box.
Then, if we set a improper $g$ for a small $\eta$ or a small $b$,
the height of particles diverge. In order to clear off such cases,
we need to control $g$ for several $\eta$ and $b$. By the simulation,
we found the fact that $CM$ have almost same values independent of $a$ with
$a<<b$ for a fixed set of $g$, $\eta$, $b$ and $V$.
Hereafter, we fix $V$ and determine the gravity $g$ for a given set
$(\eta,b)$ independent of $a$. In this paper, we set $V=2.0$ and $g(\eta,b)$ with
which $CM \sim 60d$ is realized for all cases of $a<<b$.
When we set $V=3.0$ or $V=4.0$, we can
get qualitatively same results as those of following discussions with
$V=2.0$.  

Before the discussion of the transition width at which the 1D-2D
transition takes place or the character of the transition, we discuss
the dependency of $\eta$ for the 1D states.
Figure 3 (a) shows the time averaged packing fraction profile of the $y$
direction as the function of $y-y_{CM}$ 
for several $\eta$ with $a=9d$ and $b=30d$. Here, $y_{CM}$ is the $y$
component of center of masses of particles, and we define the packing
fraction as followings. We divide the space by a lattice with the
lattice constant $d$, and we define the number of centers of particle
within each $d \times d$ square as the packing fraction in the square.  
The packing fraction of $y$ direction is given as the average of them
through the $x$ direction for each $y$.
For large $\eta$ ($\eta=0.6 , 1.0$), each profile includes a flat
region with large packing fraction near the center of mass of this
system. These flat regions mean the existence of the solid structure in 
which particles are almost completely packed.
The length of this flat region 
decreases with decreasing $\eta$, and this length becomes $0$ 
for $\eta=\eta^{*}\sim 0.38$. If $\eta<\eta^{*}$ ($\eta=0.25 , 0.35$), 
each packing fraction profile 
includes no flat region, and maximum packing fraction is smaller than 
that of $\eta>\eta^{*}$. It means that no solid structures are 
created for such small $\eta$. Thus, a transition between a state 
which includes a solid structure and the other state which include 
no solid structure occurs at the critical value $\eta=\eta^{*}$.
Similar results are obtained in following two cases, $b=20d$ and 
$b=40d$. Now we introduce following no dimensional values: $a'=a/d$,
$b'=b/d$, and $e'=1-e$ where $e=exp(\pi
\eta/(k-\eta^{2})^{\frac{1}{2}}))$, and the length of flat regions
$h(e')=h'(e')d$. ($h'(e')$ has no dimensions.)  Here, $e$ indicates
the coefficient
of restitution for head-on collisions between two particles\cite{con0}. 
The relation between two rescaled values, $e' b'$ and $\frac{h'(e')}{b'^{2}}$, is obtained as shown in
Fig.4(a). The rescaled critical point $(e' b')^{*}$ is determined
independent of the system size, and the profile of this relation fits
with
\begin{equation} 
\frac{h'(e')}{b'^{2}}=h'_{o}|e' b' - (e'b')^{*}|^{0.3}
\end{equation}
 near $(e'b')^{*} \sim 0.282$.($h'_{o}$ is constant.)
Moreover, we introduce the cluster length $\sigma(e')=\sigma'(e') d$ 
($\sigma'(e')$ has no dimensions.) which is
defined as the distance between two nearest inflection points from the
maximum point of the packing fraction profile. Thus, the relation  
between two rescaled values, $e' b'$ and $\sigma'(e') /b'$, are
obtained as shown in Fig.4 (b). The profile of this relation fits with
\begin{equation} 
\frac{\sigma'(e')}{b'} = \frac{0.0255}{(e' b')^{2}}-\frac{0.105}{e'
b'}+0.98. 
\end{equation}
We expect $\sigma'(e') /b' \to \infty$  for $e'b' \to 0$, and
$\sigma'(e') /b' \sim 1$ for $e'b' \to \infty$.
	
Now, we discuss the character of transition between the 1D state 
and the 2D state.
Here, we define this transition width $a^{*}$ as the maximum width that
the 2D state cannot be observed in the system. Figure 5 
shows the typical probability distributions of $L(t)$ for several
width $a$ around $a=\sigma(e')$. Here, (a) indicates for the case
$e'<e'^{*}$ ($\eta<\eta^{*}$) and (b) indicates for the case $e'>e'^{*}$
($\eta>\eta^{*}$)both for
$b=20d$. In Fig.5 (a), each profile of probability distribution of $L(t)$
includes only one peak. The position of peak is at 
$L(t)=0$ for  $a<\sigma(e')$. For $a>\sigma(e')$, however, the
peak appears at $L(t)>0$ and this peak moves with the width of the system.
Thus the continuous transition between the 1D state and the
2D state appears for the case $e'<e'^{*}$ ($\eta<\eta^{*}$), and the
transition width is given as $a^{*} \sim \sigma(e')$.
In Fig.5 (b), on the contrary, probability distributions of $L(t)$ for
$a$, which is a little larger than $\sigma(e')$, include two peaks
at $L(t)=0$ and $L(t)>0$. This means that two locally stable states, one
is the 1D state and the other is
the 2D state, coexist and they appear periodically for the
case of $e'>e'^{*}$ ($\eta>\eta^{*}$). In such cases, the system
includes a solid structure when 1D state is realized. In
this solid structure, the friction between particles are strong
because particles are densely packed. Hence, the solid structure 
is break-proof, and this originates the stability of the 1D 
state for $a>\sigma(e')$.
Figure 6 shows
the semi-log scale profiles of probability distributions of $L(t)$ for
respectively $a<\sigma(e')$, $a \sim \sigma(e')$ and $a>\sigma(e')$.
When $a'<\sigma'(e')$, the profile is proportion to $exp(-\gamma_{0}
L^{2})$, which means that fluctuations of $L(t)$ are so
small that they can be neglected. However, a profile which proportion
to  $exp(-\gamma_{1} L)$ is obtained when $a$ is close to $\sigma(e')$,
which means that the fluctuation from $L(t)=0$ become large. Moreover, the
profile includes two peaks at $L(t)=0$ and $L(t)>0$ when
$a \geq \sigma(e')$. Then, the transition width $a^{*}$ is regarded as
$a^{*} \sim \sigma(e')$ for also $e'>e'^{*}$ ($\eta>\eta^{*}$).
These results mean following two facts. I) The critical width which
the 2D state can appear, is
equal to the cluster length. II) The magnitude of dissipation
separates the type of the transition between the 1D state or 
the 2D state. We obtain the similar results for the case $b'=30$.
Then, by using rescaled parameter $e' b'$ and $a'/b'$, the phase 
diagram is obtained as shown in Fig.7.

In this paper, we simulated strongly excited granular materials in a 
2 dimensional pipe.
When the width of the system is small, a characteristic state, in which 
the effective dimension of particles' dynamics is regarded as 1,
appears. We named this state as the 1D state. However, when the width of
the system is larger than a 
critical width, the actual 2 dimensional state in which convection
appears. We named this state as the 2D state.
Moreover, we found that the critical width is equal to the length
of the cluster which appears in the 1D state.
The character of the 1D state depends on the magnitude of dissipation
as followings. When the magnitude of
dissipation is larger than the critical value, the system include 
solid structure. On the contrary, such solid structure disappears
when the magnitude of dissipation is smaller than the critical
value. According to the differences of the magnitude of
dissipation, following two types of behaviors appears in the system
near the critical width. When the magnitude of dissipation is
smaller than the critical value, the continuous transition between the
1D state and the 2D state appears. On the
contrary, when the magnitude of dissipation is larger than the
critical value, two meta-stable states,
the 1D state and the 2D state, appear periodically for a
little over the critical width.
By another simulation, the existence of such a critical magnitude of 
dissipation is reported\cite{fh7,fh8}. The analytical derivation of this 
critical value is one of the most important issue for the research of
granular materials. Moreover, simulations of more highly excited
systems, larger systems, and analytical study of the critical width for 
several magnitude of dissipation are important future issues.

The author is grateful to H.Nishimori for useful
discussions. This research was supported in part by the Ibaraki
University SVBL and Grant-in-Aid for JSPS Felows 10376.

\newpage

\begin{figure}[h]
\caption[]{Illustration of 2D pipe and typical snapshots respectively, (a)
$a<<b$ with $\eta>\eta_{*}$, (b) $a>>b$ with $\eta>\eta_{*}$, (c)
$a<<b$ with $\eta<\eta_{*}$, and (d) $a>>b$ with $\eta<\eta_{*}$.}
\end{figure}

\begin{figure}[h]
\caption[]{Probability distributions of $L(t)$ respectively (a)
$a<<b$, and (b) $a>>b$.}
\end{figure}

\begin{figure}[h]
\caption[]{Packing fraction profile for $y$ direction respectively
$\eta=0.25$,$ 0.35$,$ 0.6$, and $1.0$.}
\end{figure}

\begin{figure}[h]
\caption[]{Relation between two rescaled values, (a) $e' b'$ and
$\frac{h'(e')}{b'^{2}}$, and (b) $e' b'$ and $\sigma'(e') /b'$}
\end{figure}

\begin{figure}[h]
\caption[]{Typical probability distributions of $L(t)$ of (a)
$e'<e'^{*}$ ($\eta<\eta^{*}$) and (b) $e'>e'^{*}$ ($\eta>\eta^{*}$) for several width around a
critical value $a=\sigma(e')$.}
\end{figure}

\begin{figure}[h]
\caption[]{Semi-log scale profiles of probability distributions of $L(t)$ for
respectively $a<\sigma(e')$, $a$ is a little smaller than
$\sigma(e')$ and $a \geq \sigma(e')$ with $e'\sim0.247$ $(\eta=1.0)$}
\end{figure}

\begin{figure}[h]
\caption[]{Phase diagram of the effective dimension of the system
which depends on $e' b'$ and $a'/b'$}
\end{figure}


\begin{thebibliography}{999}
\bibitem{setu}
Similar simplified situations have been investigated to discuss the
statistical and hydro-dynamical properties of granular
materials\cite{th1,th2,th3}.
\bibitem{keta} The elastic constant $k$ and the viscosity coefficient
$\eta$ are related with the coefficient of restitution $e$ and the
collision time $t_{col}$, time period during collision [2].
\end{thebibliography}
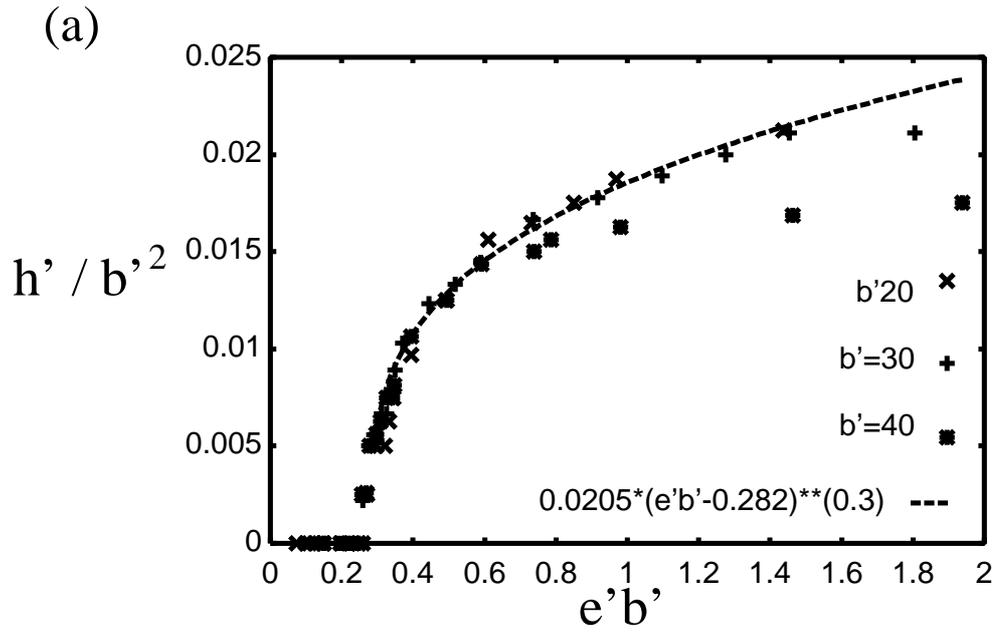
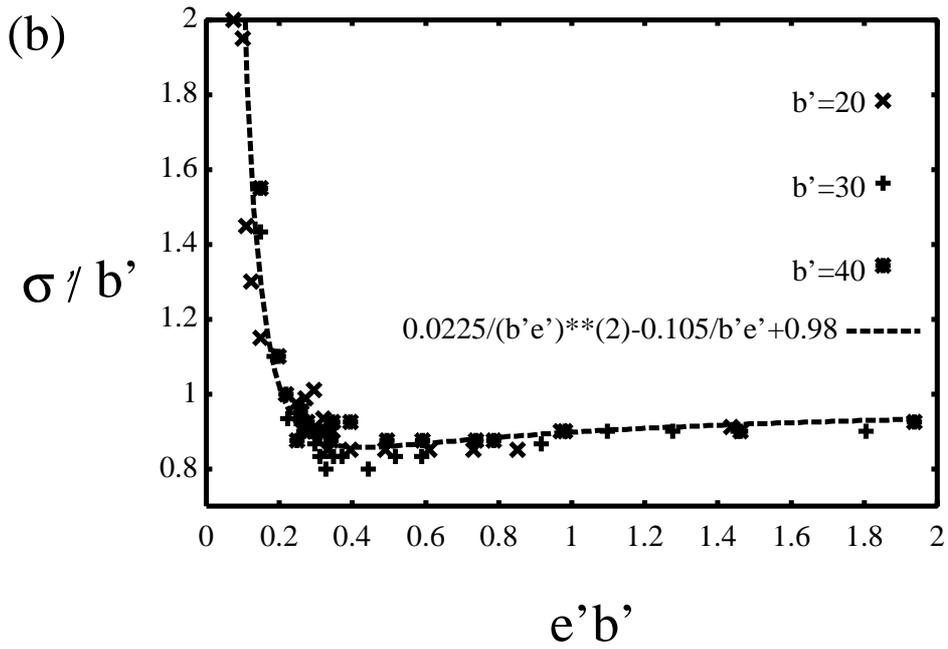
\end{document}